\documentclass[preprint]{aastex62} 

\begin{document}

\title{A Deep and Wide Twilight Survey for Asteroids Interior to Earth and Venus}

\author[0000-0003-3145-8682]{Scott S. Sheppard} 
\affil{Earth and Planets Laboratory, Carnegie Institution for Science, 5241 Broad Branch Rd. NW, Washington, DC 20015, USA, ssheppard@carnegiescience.edu}

\author[0000-0003-0773-1888]{David J. Tholen} 
\affil{Institute for Astronomy, University of Hawai'i, Honolulu, HI 96822, USA}

\author[0000-0002-5667-9337]{Petr Pokorn\'{y}} 
\affiliation{Department of Physics, The Catholic University of America, Washington, DC 20064, USA}
\affiliation{Astrophysics Science Division, NASA Goddard Space Flight Center, Greenbelt, MD 20771}
\affiliation{Center for Research and Exploration in Space Science and Technology, NASA/GSFC, Greenbelt, MD 20771}

\author[0000-0001-7895-8209]{Marco Micheli}
\affil{ESA NEO Coordination Centre, Largo Galileo Galilei, 1, 00044 Frascati (RM), Italy}

\author{Ian Dell'Antonio}
\affil{Physics Department, Brown University, Box 1843, Providence, RI 02912, USA}

\author{Shenming Fu} 
\affil{Physics Department, Brown University, Box 1843, Providence, RI 02912, USA}
\affil{NSF’s National Optical-Infrared Astronomy Research Laboratory, 950 North Cherry Avenue, Tucson, AZ 85719, USA}

\author[0000-0001-9859-0894]{Chadwick A. Trujillo}
\affil{Department of Astronomy and Planetary Science, Northern Arizona University, Flagstaff, AZ 86011, USA}

\author[0000-0002-5382-2898]{Rachael Beaton} 
\affil{Department of Astrophysical Sciences, Princeton University, 4 Ivy Lane, Princeton, NJ 08544, USA}
\affil{The Observatories of the Carnegie Institution for Science, 813 Santa Barbara St., Pasadena, CA~91101}

\author[0000-0002-5382-2898]{Scott Carlsten} 
\affil{Department of Astrophysical Sciences, 4 Ivy Lane, Princeton University, Princeton, NJ 08544, USA}

\author[0000-0001-8251-933X]{Alex Drlica-Wagner} 
\affil{Fermi National Accelerator Laboratory, P.O.\ Box 500, Batavia, IL 60510, USA}
\affil{Department of Astronomy and Astrophysics, University of Chicago, Chicago, IL 60637, USA}

\author[0000-0002-9144-7726]{Clara Mart\'inez-V\'azquez} 
\affil{Gemini Observatory, NSF's NOIRLab, 670 N. A'ohoku Place, Hilo, HI 96720, USA }
\affil{Cerro Tololo Inter-American Observatory/NSF’s NOIRLab, Casilla 603, La Serena, Chile}

\author[0000-0003-3519-4004]{Sidney Mau} 
\affil{Department of Physics, Stanford University, 382 Via Pueblo Mall, Stanford, CA 94305, USA}

\author[0000-0002-0143-9440]{Toni Santana-Ros} 
\affil{Departamento de F\'isica, Ingenier\'ia de Sistemas y Teor\'ia de la Se\~nal, Universidad de Alicante, Carr. de San Vicente del Raspeig, s/n, 03690 San Vicente del Raspeig, Alicante, Spain}
\affil{Institut de Ci\'encies del Cosmos (ICCUB), Universitat de Barcelona (IEEC-UB), Carrer de Mart\'i i Franqu\'es, 1, 08028, Barcelona, Spain}

\author[0000-0003-3402-6164]{Luidhy Santana-Silva} 
\affil{NAT-Universidade Cruzeiro do Sul / Universidade Cidade de São Paulo, Rua Galvão Bueno, 868, 01506-000, São Paulo, SP, Brazil}

\author[0000-0002-8149-1352]{Crist\'obal Sif\'on} 
\affil{Instituto de F\'isica, Pontificia Universidad Cat\'olica de Valpara\'iso, Casilla 4059, Valpara\'iso, Chile}

\author[0000-0003-3801-1496]{Sunil Simha} 
\affil{University of California at Santa Cruz, 1156 High St., Santa Cruz, CA, 95064 USA}

\author[0000-0002-1506-4248]{Audrey Thirouin} 
\affil{Lowell Observatory, 1400 W Mars Hill Road, Flagstaff, AZ 86001, USA}

\author[0000-0003-4580-3790]{David Trilling} 
\affil{Northern Arizona University, Flagstaff, AZ 86011, USA}

\author[0000-0003-4341-6172]{A.~Katherina~Vivas} 
\affil{Cerro Tololo Inter-American Observatory/NSF’s NOIRLab, Casilla 603, La Serena, Chile}

\author[0000-0001-6455-9135]{Alfredo Zenteno} 
\affil{Cerro Tololo Inter-American Observatory/NSF’s NOIRLab, Casilla 603, La Serena, Chile}







\begin{abstract}  

We are conducting a survey using twilight time on the Dark Energy
Camera with the Blanco 4m telescope in Chile to look for objects
interior to Earth's and Venus' orbits. To date we have discovered two
rare Atira/Apohele asteroids, 2021 LJ4 and 2021 PH27, which have
orbits completely interior to Earth's orbit.  We also discovered one
new Apollo type Near Earth Object (NEO) that crosses Earth's orbit,
2022 AP7. Two of the discoveries have diameters $\gtrsim 1$ km. 2022
AP7 is likely the largest Potentially Hazardous Asteroid (PHA)
discovered in about eight years.  To date we have covered 624 square
degrees of sky near to and interior to the orbit of Venus.  The
average images go to 21.3 mags in the r-band, with the best images
near 22nd mag. Our new discovery 2021 PH27 has the smallest semi-major
axis known for an asteroid, 0.4617 au, and the largest general
relativistic effects (53 arcseconds/century) known for any body in the
Solar System.  The survey has detected $\sim 15\%$ of all known Atira
NEOs. We put strong constraints on any stable population of Venus
co-orbital resonance objects existing, as well as the Atira and Vatira
asteroid classes.  These interior asteroid populations are important
to complete the census of asteroids near Earth, including some of the
most likely Earth impactors that cannot easily be discovered in other
surveys. Comparing the actual population of asteroids found interior
to Earth and Venus with those predicted to exist by extrapolating from
the known population exterior to Earth is important to better
understand the origin, composition and structure of the NEO
population.

\end{abstract}

\keywords{Small Solar System bodies, Asteroid, Minor Planets, Near-Earth objects, Atira group}

\section{Introduction}

The small body population in our Solar System interior to Earth's and
Venus' orbits has not been extensively explored to date because of the
difficulty in observing near the glare of the Sun. Twilight
observations near Venus' orbit from ground-based Earth telescopes occur
at a very high airmass, with a bright sky background, poorer than
average seeing and with the objects of interest being at high phase
angles during a short observational window. These factors create well
over a magnitude of signal-to-noise loss when compared to observations
overhead at night.  This region of space is important for
understanding the distribution of Near Earth Objects (NEOs) and
objects that may have stable orbits in resonance with Venus.

One way to better estimate the true number of small NEOs is to include
more objects found interior to the Earth's orbit in the population
calculations, increasing the completion of the orbital NEO types
(Granvik et al. 2018; Harris \& Chodas 2021). Currently the population
models are biased towards NEOs found exterior to the Earth's orbit as
they are the easiest to find observationally.  Only about 25 asteroids
are known that have orbits completely interior to the Earth's
orbit and have well-determined orbits (called Atira or Apohele
asteroids).  This is compared to the thousands of known NEOs with
orbits that cross Earth's orbit such as the Aten and Apollo NEOs with
semi-major axes interior and exterior to Earth, respectively
(Morbidelli et al. 2020; Schunova-Lilly et al. 2017; Mainzer et
al. 2014).

Understanding the relative populations of objects with orbits interior
to Earth versus exterior to Earth will help us understand how objects
are transported throughout the inner Solar System (Strom et al. 2015).
In addition, disruption processes of NEOs, such as fragmenting from
thermal stresses, breaking up from rotational excitation or tidal
disruption from passing near the Sun or planets can be analyzed by
comparing the predicted number of NEOs found interior to Earth versus
exterior to Earth and correlating this with the composition or type of
objects found (Granvik et al. 2016).  Such correlations and analysis
will further yield insights into the structure and strength of the
different types of NEO classes.  Various internal stresses on the NEOs
such as tidal deformation and solar heating will cause more fragile
asteroid compositions and structures to erode or break-up as they
approach the planets and Sun (Li \& Jewitt 2013; Ye \& Granvik
2019).

\begin{figure}
\epsscale{0.4}
\centerline{\includegraphics[angle=0,totalheight=0.4\textheight]{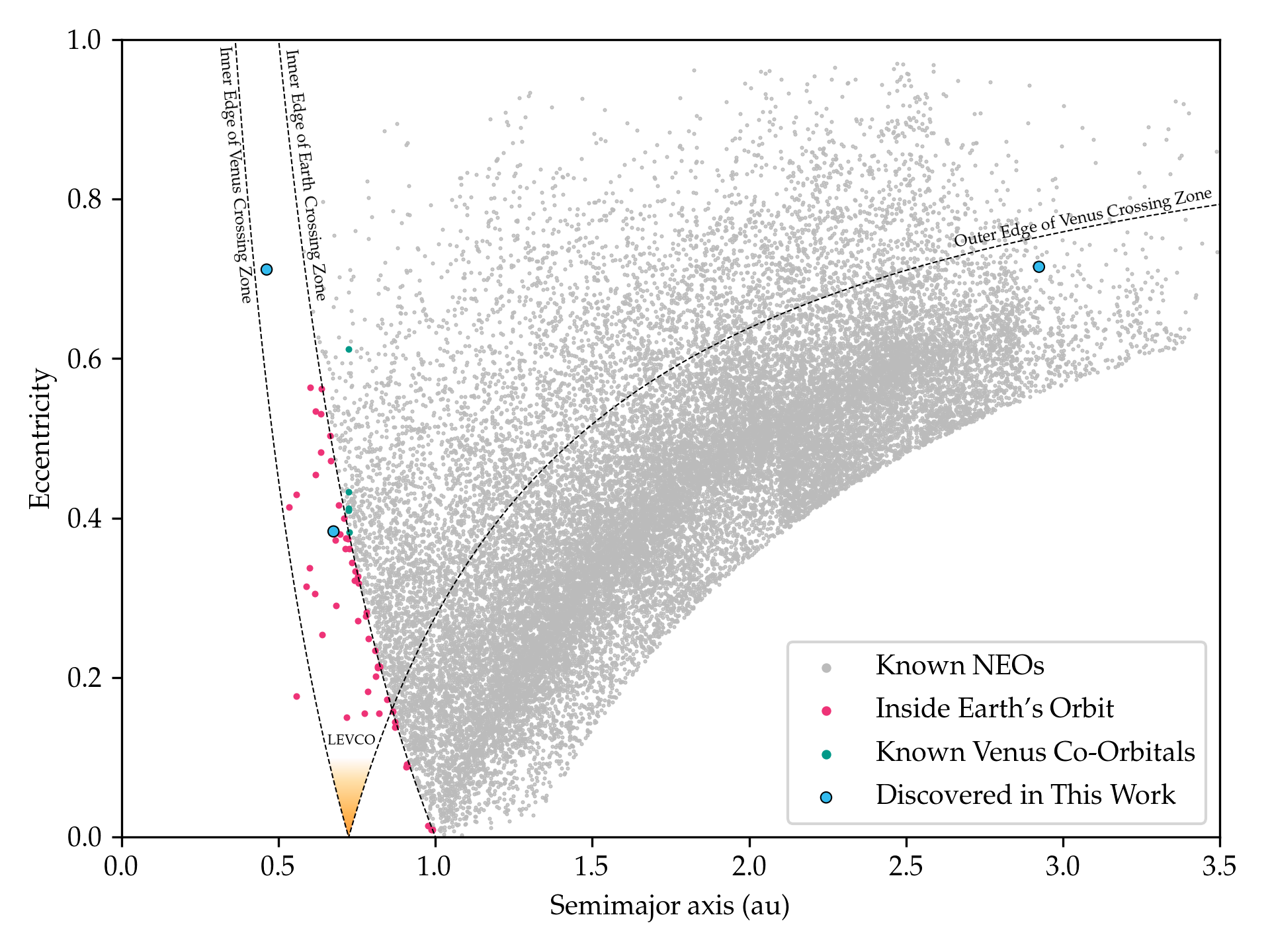}}
\caption{Known Near Earth Objects (NEOs) plotted with semi-major axis
  versus eccentricity.  The new discoveries from this survey are shown
  by big blue circles. The yellow area shows the region near Venus'
  orbit that dynamcially stable Low Eccentricity Venus Co-Orbitals
  (LEVCO) would be expected to be found, where no asteroid is known.
  All known asteroids with semi-major axes similar to Venus are on
  dynamically unstable highly eccentric orbits that cross Earth's
  orbit (small green circles).  Atira type object's orbits remain
  interior to the Earth's orbit, shown as red circles to the left of
  the vertical dashed line starting at 1 au, which is the inner edge
  of the Earth crossing zone.
\label{fig:ae_plot} }
\end{figure}

\begin{figure}
\epsscale{0.4}
\centerline{\includegraphics[angle=0,totalheight=0.4\textheight]{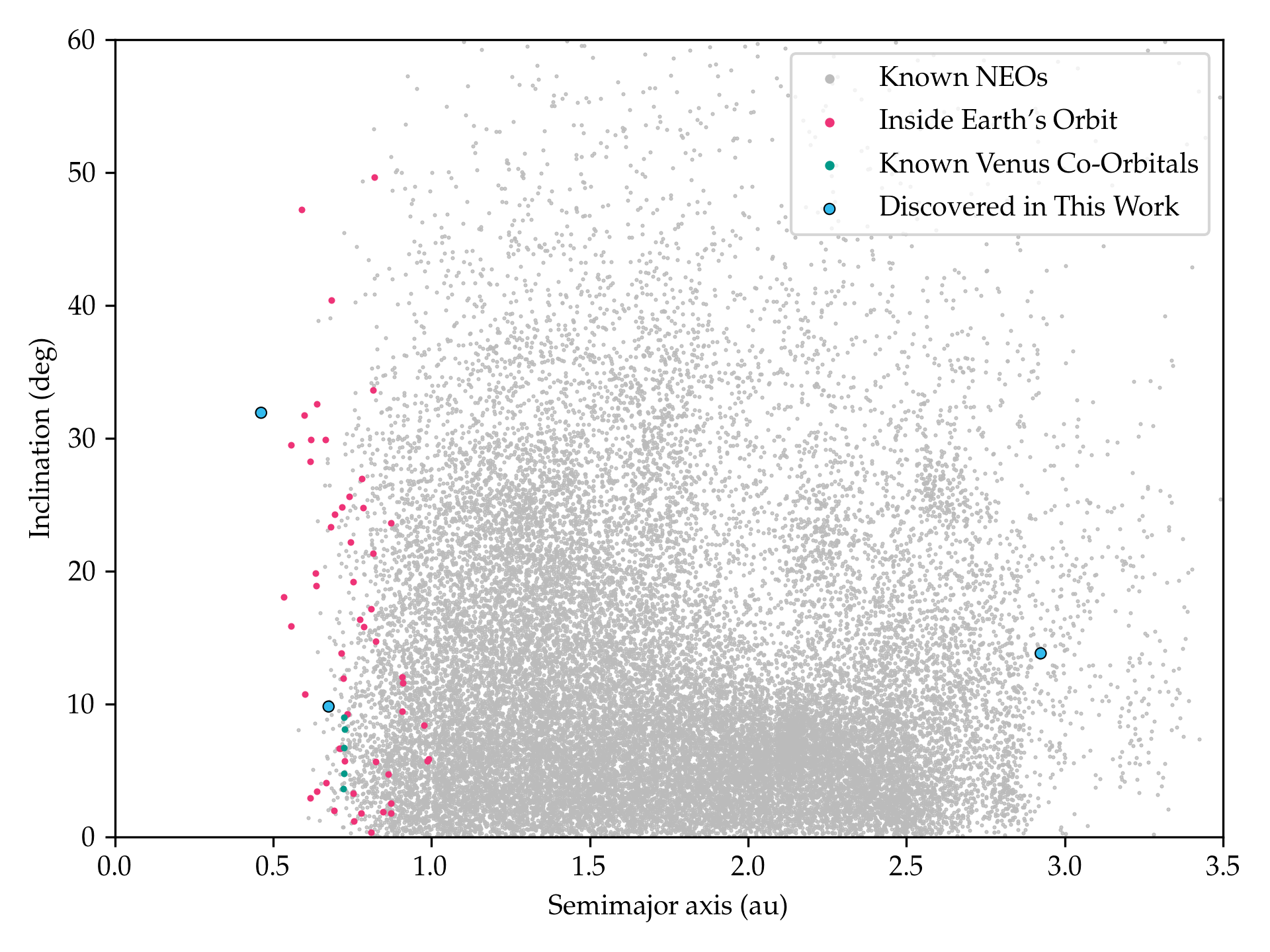}}
\caption{Same as Figure~\ref{fig:ae_plot} but using inclination.
\label{fig:ai_plot} }
\end{figure}

In addition to NEOs, which are strongly influenced by the Earth, some
objects found interior to Earth's orbit may exhibit strong influences
from Venus. There are a few known well observed asteroids that have
orbital periods similar to Venus': (322756) 2001 CK32, (524522) 2002
VE68, 2012 XE133, 2013 ND15 and 2015 WZ12 as well as some more
recently discovered NEOs with semi-major axes near Venus' between
0.722 and 0.725 au: 2020 BT2, 2020 CL1, 2020 QU5, 2021 XA1, 2021 XO3,
2022 BL5 and 2022 CD (see Figures~\ref{fig:ae_plot}
and~\ref{fig:ai_plot}).  All of these near Venus co-orbital asteroids
are dynamically unstable on million-year timescales since they have
high eccentricity orbits that cross the orbit of the Earth (Mikkola et
al. 2004; de la Fuente Marcos \& de la Fuente Marcos
2012,2013,2014,2017).  The same situation has been found for the known
co-orbitals of Earth (Brasser et al. 2004; Connors et al. 2011; Hui et
al. 2021; Santana-Ros et al. 2022).  The population of low
eccentricity Venus and Earth co-orbitals that are stable resonant
objects may be small due to perturbations from the terrestrial planets
or non gravitational effects (Morais \& Morbidelli 2006; Malhotra
2019; Pokorny \& Kuchner 2021). There are no known satellites of
Venus, likely because the Hill Sphere of Venus is mostly dynamically
unstable to long term satellites (Sheppard and Trujillo 2009).  To
date, the population of 1 km and smaller objects near Venus' orbit is
relatively unconstrained observationally as most surveys in this
region are only sensitive to larger objects.

Data from the two space missions HELIOS and STEREO indicate the
presence of a narrow ring of dust in Venus' orbit (Leinert \& Moster
2007; Jones et al. 2013).  This Venus co-orbital dust ring could come
from a population of low eccentricity ($e<0.3$) stable Venus resonant
co-orbital objects (Pokorny \& Kuchner 2019).  These authors also
found that about $8\%$ of an initial population of Venus resonant
objects remain stable for the age of the solar system.  Thus, these
putative low eccentricity Venus resonant co-orbitals could be the
leftover remnants of the planetesimals that formed near Venus, unlike
most NEOs and high eccentricity Venus co-orbitals that are believed to
have recently escaped from the much further out main asteroid belt
(Granvik et al. 2017).  This theorized population of low eccentricity
Venus resonant objects has yet to be observationally ruled out since
most surveys to date have not covered the large areas of sky to the
faint depths needed to find such a population of low eccentricity
objects interior to Earth.  This population could be too faint for
small class telescopes to efficiently detect in the glare of the Sun
as objects less than about 1 km in size would mostly be fainter than
about 20th magnitude (Morbidelli et al. 2020; Masiero et al. 2020).

Surveys that have covered significant sky looking for asteroids near
Venus have been small class telescopes (Myhrvold 2016). Recently the
48 inch telescope Zwicky Transient Facility (ZTF: Ye et al. 2020)
survey has done the most extensive search for asteroids near Venus.
ZTF found the first ever asteroid that has an orbit entirely interior
to Venus, (594913) 'Ayl\'{o}'chaxnim (2020 AV2) (Bolin et al. 2020),
though the orbit is only stable for a few million years (Greenstreet
2020; de la Fuente Marcos \& de la Fuente Marcos 2020; Bolin et
al. 2021).  This telescope has covered several thousand square degrees
of sky interior to Earth's orbit, but is limited in its discovery
ability of the smaller objects as it uses a 1.2 meter telescope (Ye et
al. 2020). Based on 'Ayl\'{o}'chaxnim's absolute magnitude, it is
likely a relatively large asteroid of $\sim 1.5$ km in size.

We performed a pilot survey in September 2019 to determine if a search
for objects near Venus' orbit is feasible with the medium class Blanco
4 meter telescope using the large field of view Dark Energy Camera
(DECam), which is the largest sky area camera on a 4 meter or larger
telescope.  We searched about 35 square degrees of sky some 40 to 50
degrees in elongation away from the Sun near the end of nautical and
start of astronomical twilight time in the evening.  The images went
to over 21st magnitude in the r-band and we were able to put some
moderate limits on the size of any stable Venus resonant co-orbital
population (Pokorny et al. 2020).  Here we discuss an additional 589
square degrees of twilight sky near to and interior to Venus' orbit,
yielding a total of 624 sq deg searched in the survey to date.

\section{Observations}

The twilight survey for objects near to and interior to Venus' orbit
uses the DECam on the Cerro-Tololo Inter-American Observatory (CTIO) 4
meter Blanco telescope in Chile. DECam has 61 working science CCDs
arranged in a circular type pattern at the prime focus of the
telescope.  Each CCD has $2048 \times 4096$ pixels with a pixel scale
of about 0.264 arcseconds, yielding a field-of-view of about 2.7
square degrees per image (Flaugher et al. 2015). Images were
calibrated by subtracting a median bias image and divided by a median
dome flat.

\begin{figure}
\epsscale{0.4}
\centerline{\includegraphics[angle=180,totalheight=0.5\textheight]{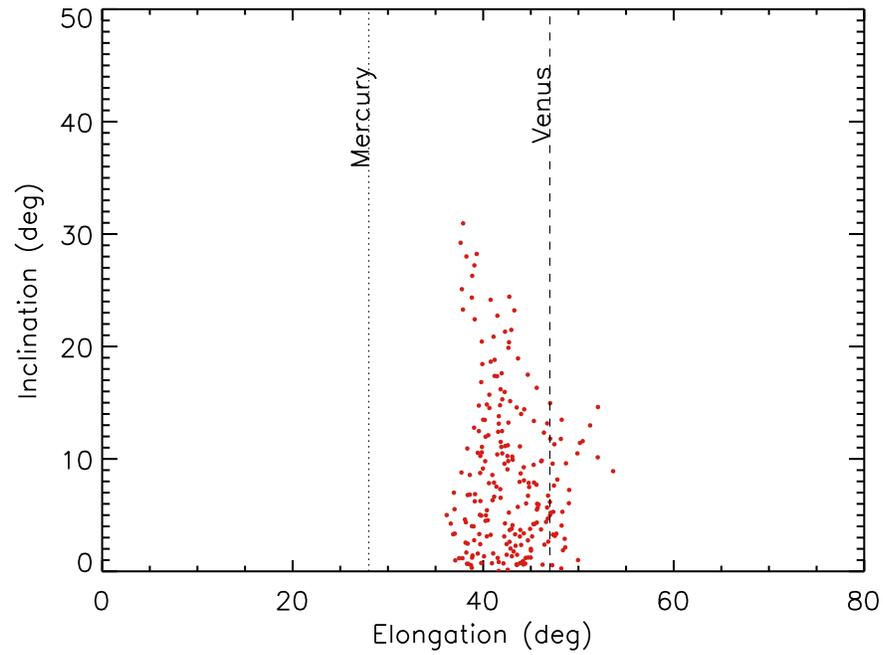}}
\caption{The twilight survey fields taken with DECam showing their
  elongation position from the Sun and inclination from the
  ecliptic. The maximum elongation from the Sun of Mercury and Venus
  are shown by vertical dotted and dashed lines respectively.
\label{fig:TwilightElong_Incl} }
\end{figure}

Observations were obtained just after nautical twilight ends in the
evening and just before nautical twilight begins in the morning.  The
Sun is usually between about -15 and -12 degrees below the horizon
when the images are taken, giving about 10 minutes of observation time
each twilight. Images were generally taken at airmasses between about
2.3 and 2.5, near the 23 degree elevation limit of the telescope.
Most fields were observed near to or interior to Venus' orbit when
projected onto the sky (Figure~\ref{fig:TwilightElong_Incl}). The
r-band filter with exposure times of 28 seconds were used. With
simultaneous readout and offsetting the telescope to the next position
a few degrees away, it takes just under 1 minute between the start of
two successive images. Two images of each field are taken separated by
about 2 to 4 minutes. The images usually only had a background of a
few thousand counts, while the detector does not saturate until
several tens of thousands of counts.

The depth of each image mostly depends on the seeing and the sky
brightness. The signal-to-noise ratio of asteroids in the twilight
images is generally worse than when observed exterior to Earth's orbit
because of the high phase angles of the asteroids, bright background
sky and high airmass.  In the twilight images taken near the Sun, the
main belt asteriods are much farther away in our fields than when
imaged in darker skies exterior to the Earth's orbit. Thus, because of
the relatively large distances of the main belt asteroids, our survey
will not usually find unknown main belt asteroids.

\begin{figure}
\epsscale{0.4}
\centerline{\includegraphics[angle=0,totalheight=0.4\textheight]{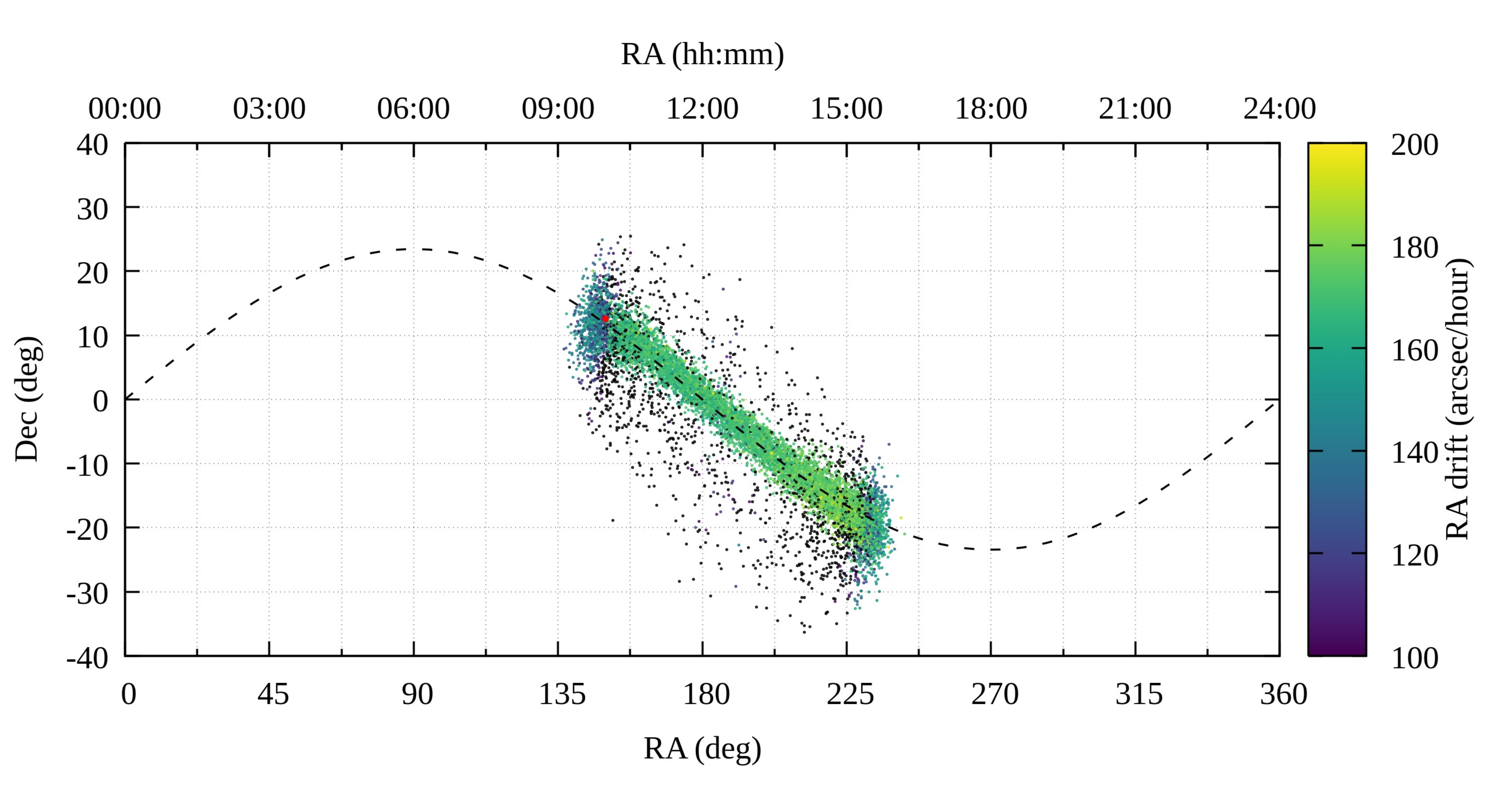}}
\caption{The Right Ascension and Declination of a hypothetical
  population of low eccentricity, low inclination, stable Venus
  co-orbital asteroids (Pokorny \& Kuchner 2019) as they would appear
  on 2020 September 30 at 23:30 UT.  Objects near Venus' orbit well
  interior to the Earth's orbit would appear to move faster than about
  100 arcseconds per hour, distinguishing them from more distant main
  belt asteroids that tend to move slower than 80 arcseconds per
  hour. The red dot shows Venus' location.
\label{fig:VenusCoorbitalpaper} }
\end{figure}

An object that is interior to the Earth's orbit will generally have an
apparent motion of over 100 arcseconds per hour, with objects near
Venus' orbit expected to show motions near 150 arcseconds per hour
(Figure~\ref{fig:VenusCoorbitalpaper}).  The exposure time is limited
by the fast motion of the objects, as an object with an apparent
motion of 150 arcseconds per hour will start to show significant
trailing after about 30 seconds.  The survey used 28 seconds for the
exposure time to limit any signal-to-noise trailing losses of the
asteroids.  The moving object search algorithm was the same as that
used to find outer solar system objects (Trujillo et al. 2001;
Sheppard et al. 2019) and it was set to flag any objects that were
found to have an apparent motion faster than 75 arcseconds per hour.
This allows us to easily identify the objects of interest that could
be NEOs or Venus resonant co-orbitals while rejecting the many objects
within the main asteroid belt and beyond, as their apparent motion is
less than about 80 arcseconds per hour.  Since some main belt
asteroids have apparent motions faster than our 75 arcsecond per hour
lower limit, we used the known main belt asteroids we did detect
moving faster than 75 arcseconds per hour to help characterize our
survey's detection efficiency.

We determined the limiting magnitude or depth to find moving objects
in our survey fields based on two techniques. First, we determined to
what magnitude we detected known main belt asteroids as well as known
NEOs.  Second, we implanted artificial moving objects into some fields
to determine our moving object detection efficiency. The techniques
are complementary and agreed to within a tenth of a magnitude of each
other.  The known main belt asteroids had their V-band Minor Planet
Center magnitudes converted to r-band magnitudes using the simple
conversion $V-r=0.2$ mags (Smith et al. 2002; Sheppard 2012; Sergeyev
\& Carry 2021). Artificial objects were placed in some of the fields
ranging from 19th to 23rd magnitude in the r-band with apparent
motions as expected for NEOs between 80 and 170 arcseconds per hour.
Because of the short time-base of only $\sim 4$ minutes between images
and the short integration time of 28 seconds, we found no significant
differences in our detection of objects with the fastest or slowest
apparent motions.

\begin{figure}
\epsscale{0.4}
\centerline{\includegraphics[angle=180,totalheight=0.5\textheight]{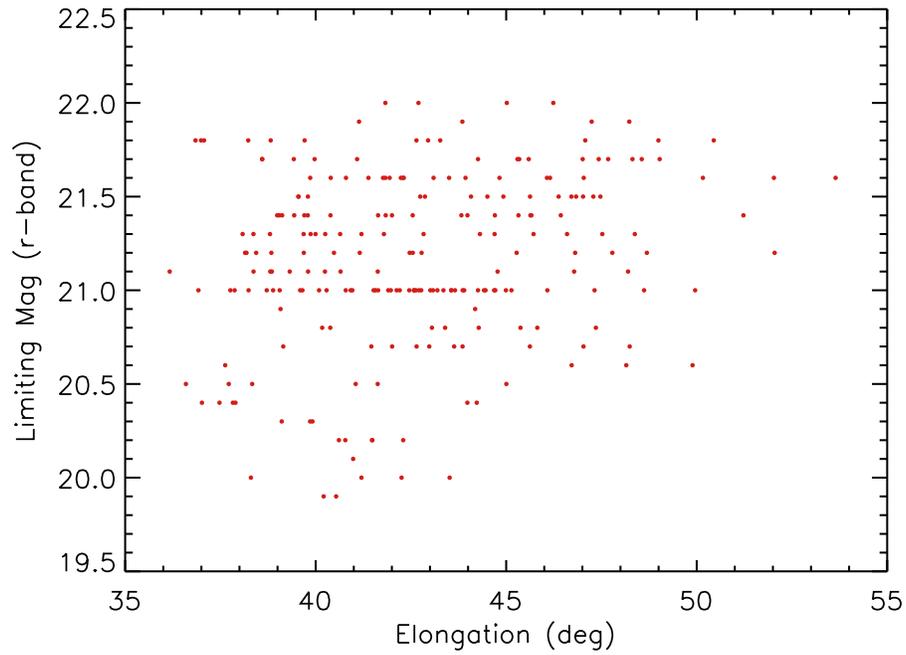}}
\caption{The limiting magnitude of each twilight field versus its
  elongation distance from the Sun.  Most of the fields had a limiting
  magnitude above 21 mags in the r-band, with the best fields reaching
  near 22nd magnitude.
\label{fig:Limitingmag_Elong}}
\end{figure}

\begin{figure}
\epsscale{0.4}
\centerline{\includegraphics[angle=180,totalheight=0.5\textheight]{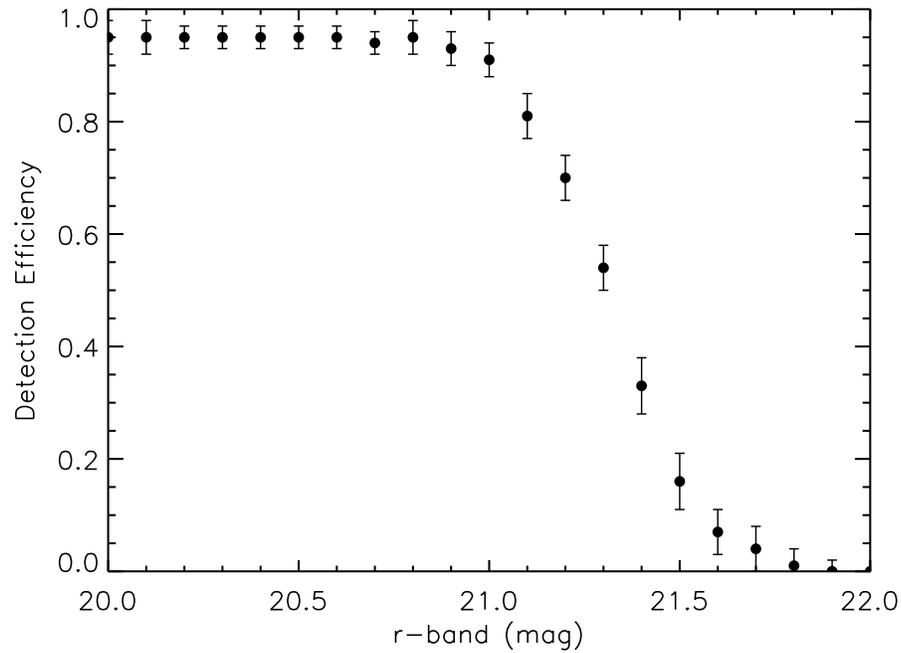}}
\caption{The moving object detection efficiency for the survey for a typical field with 1.4 arcsecond seeing. The typical field limiting magnitude was around 21.3 mags in the r-band. Table 1 and Figure~\ref{fig:Limitingmag_Elong} shows the limiting magnitude of each field of the DECam twilight survey.
\label{fig:Efficiencytwilight2022}}
\end{figure}

We find that for the average field with a seeing of about 1.4
arcseconds, we detected about 50\% of artificially implanted objects
for $m_{r}\sim 21.3$ mag, which is what we take as the overall
limiting magnitude of the survey, though different nights had
different limiting magnitudes as shown in Table 1 and
Figure~\ref{fig:Limitingmag_Elong}, with some of the best seeing
fields reaching near 22nd magnitude in depth. The detection efficiency
curve has a similar shape each night as shown in
Figure~\ref{fig:Efficiencytwilight2022}, just shifted in $m_{r}$ to
the value as shown in Table 1 mostly depending on the
seeing. Figure~\ref{fig:Limitingmag_Elong} shows the limiting
magnitude of each field versus the elongation of the field from the
Sun. The survey was able to detect asteroids near Venus' orbit of only
a few hundred meters in size assuming moderate albedos like S-type
asteroids (Pokorny et al. 2020).

From Jedicke et al. (2016) we use Equation (15) to fit the detection
efficiency curve shown in Figure~\ref{fig:Efficiencytwilight2022}:
$\epsilon (m_{r}) = \epsilon_{o}[1+\exp(m_{r}-r_{50\%}/r_{width})]^{-1}$, where $r_{50\%}$ is the
magnitude in the r-band where the efficiency is $50\%$,
$\epsilon _{o}$ is the maximum survey efficiency (i.e., the brightest
object efficiency), and $r_{width}$ is the magnitude range in $m_{r}$
that the efficiency decreases from near 75\% to 25\%.  We find
$\epsilon _{o} = 0.95\pm0.01$, $r_{50\%} = 21.3$ and $r_{width} =
0.19$.

As the survey was active every few nights, most of the asteroid
recovery of new discoveries was done using DECam. D. Tholen also
recovered new discoveries with the University of Hawaii 88-inch
telescope and the Canada-France-Hawaii 3.6 meter telescope (CFHT). The
6.5 meter Baade Magellan telescope at Las Campanas in Chile was also
used in recovery observations as were the 1 meter global network of
telescopes operated by Las Cumbres Observatory, mainly through the
European Space Agency's (ESA) Planetary Defense Office (PDO).

\section{Results}



Using DECam, we covered 624 square degrees of sky near to and interior
to Venus' orbit. Three new NEOs were discovered, with two likely being
about 1 km or larger in size (Table 2).  In addition, we serendipitously
detected several known NEOs moving well over 90 arcseconds per hour,
including Atira, Aten, Amor, and Apollo type NEOs, with several being
Potentially Hazardous Asteroids (PHAs).

\subsection{2021 PH27: Atira Type}

Asteroid 2021 PH27 was discovered by S. Sheppard on UT 2021 August 13
in twilight images taken near the Sun by the Local Volume Complete
Cluster Survey (LoVoCCS: Dell'Antonio 2020; Fu et al. 2022) in
collaboration with the DECam twilight asteroid survey (Sheppard et
al. 2021). The asteroid was recovered the next night on UT August 14
at both the Baade-Magellan telescope and again using DECam on the
Blanco 4 meter telescope.  As the object was relatively bright at some
19.2 mags, it could be tracked by 1 meter class telescopes.  On UT
August 15 the asteroid was observed by DECam and Magellan as well as
the Las Cumbres 1 meter telescopes in South Africa and Chile.  It was
tracked with 1 meter class telescopes for almost a month.  In
addition, pre-discovery images of the asteroid were found from DECam
archival images on UT 2017 July 16, allowing the orbit to be
well-determined. 2021 PH27 was also recovered again in March 2022.

\begin{figure}
\epsscale{0.4}
\centerline{\includegraphics[angle=180,totalheight=0.6\textheight]{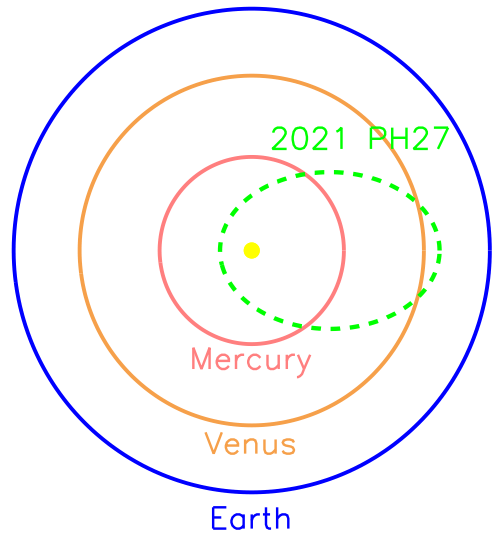}}
\caption{The orbit of newly discovered Atira asteroid 2021 PH27, which
  has the smallest semi-major axis of any known asteroid.  Though it
  orbits the Sun faster than Venus as its semi-major axis is less than
  Venus, 2021 PH27 has an aphelion exterior to Venus.  It also has a
  perihelion interior to Mercury's orbit, causing 2021 PH27 to
  experience the largest General Relativistic effects from the Sun's
  gravity of any object known in the Solar System.
\label{fig:v13aug1plan} }
\end{figure}

2021 PH27 has the smallest semi-major axis of any known asteroid and
thus the shortest orbital period for an asteroid of about 113 days
(Table 2).  Only the planet Mercury has a smaller known semi-major
axis for any object in our Solar System.  2021 PH27 has an Atira type
asteroid orbit that always keeps it inside of Earth's orbit
(Figure~\ref{fig:v13aug1plan}).  Because 2021 PH27 has a fairly
eccentric orbit, it crosses both the orbits of Mercury and Venus and
approaches the Sun to within about 0.133 au.

To understand the orbital behavior of 2021 PH27, we ran a numerical
orbital simulation using the SWIFT RMVS4 integrator described in
Levison \& Duncan (1994) and included the planets Mercury, Venus,
Earth+Moon (barycenter), Mars (barycenter), Jupiter (barycenter),
Saturn (barycenter), Uranus (barycenter) and Neptune (barycenter). The
time step of the simulations were 0.1 days.  We generated 1000 clones
using the JPL covariance matrix of 2021 PH27 from April 18, 2022.
The randomly generated points follow a 6 dimensional normal
distribution defined by the covariance matrix and the orbital elements
as described in Namouni \& Morais (2018).

\begin{figure}
\epsscale{0.4}
\centerline{\includegraphics[angle=0,totalheight=0.4\textheight]{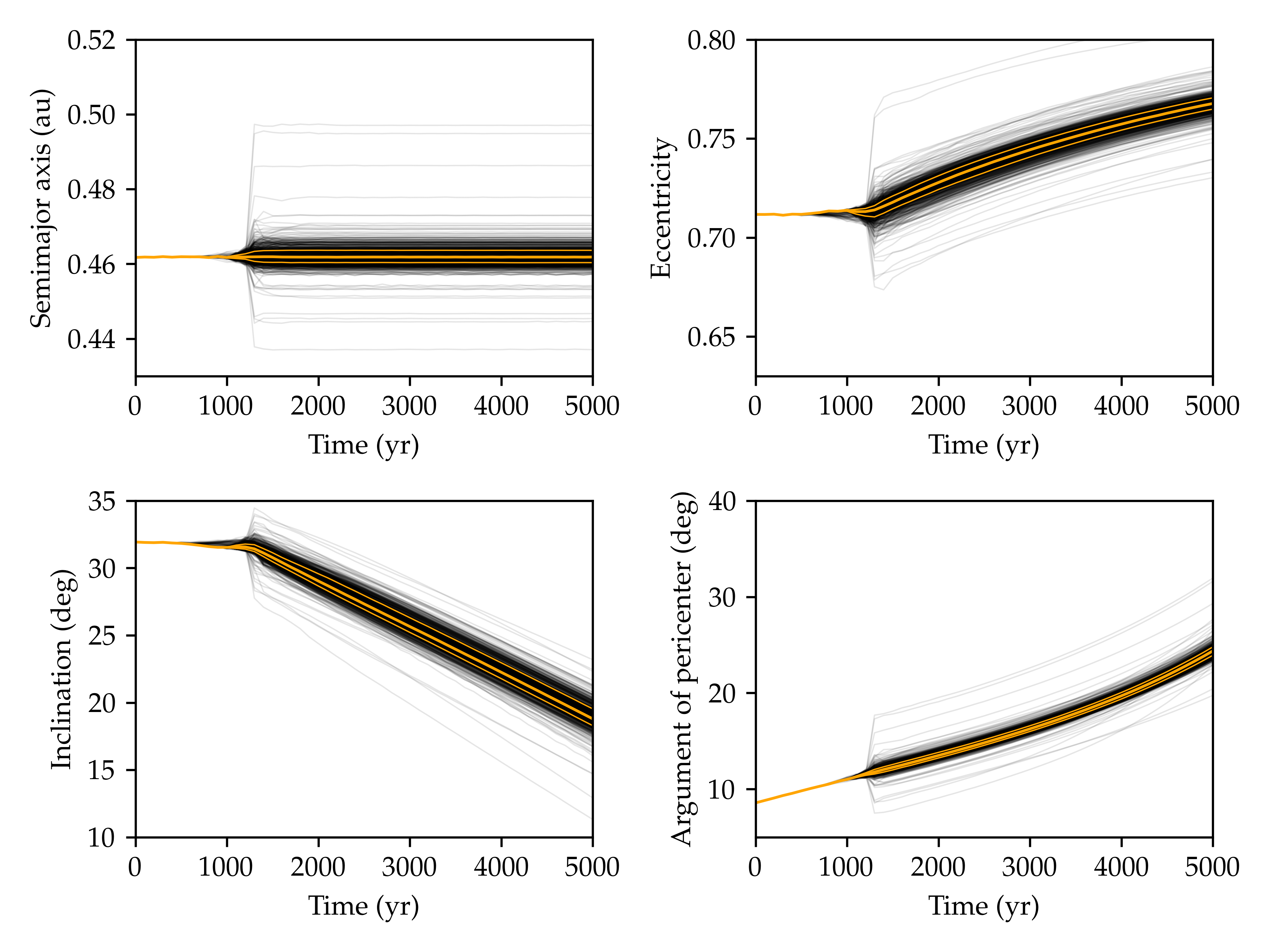}}
\caption{ The orbital stability of 2021 PH27 over a few thousand years
  time. Each grey line shows one of the 1000 clone particles of 2021
  PH27's orbital parameters.  The bolded middle yellow line shows the
  median orbital parameters of 2021 PH27 and the thin yellow lines the
  1 sigma of the median value. The orbit is dynamically unstable and
  2021 PH27 has numerous close encounters with Venus. The first close
  encounter with Venus occurs around the year 3000 AD (or 950 to 1050
  years from now or time zero). This close encounter with Venus
  creates a noticable spread in the asteroid clones of 2021 PH27, as
  most of these clones get inside of Venus' Hill Sphere during the
  encounter, modifying the orbit of 2021 PH27 depending on how close
  it gets to the planet.
\label{fig:2021PH27stability1} }
\end{figure}

\begin{figure}
\epsscale{0.4}
\centerline{\includegraphics[angle=0,totalheight=0.4\textheight]{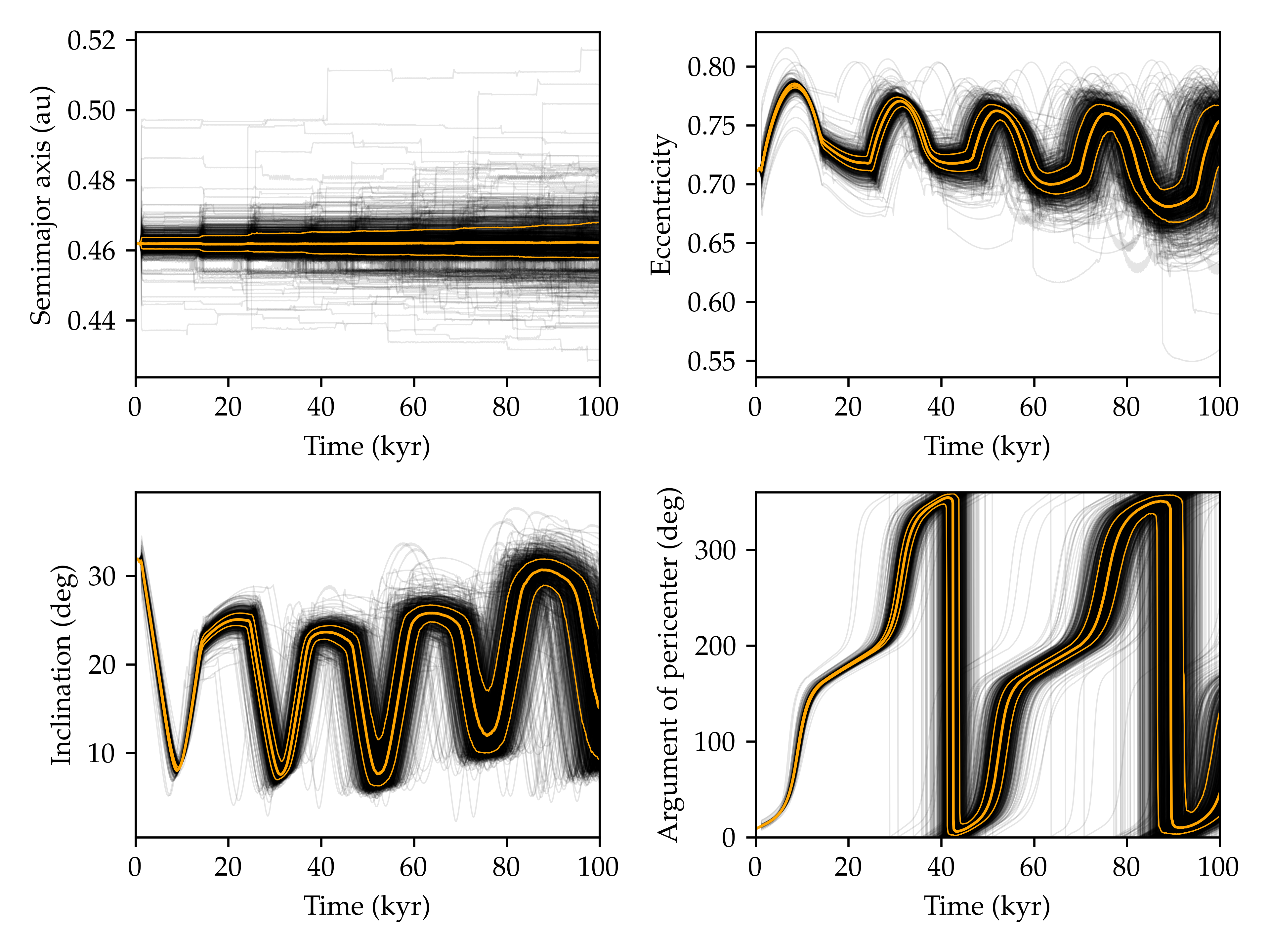}}
\caption{ Same as Figure~\ref{fig:2021PH27stability1} but now showing
  the numerical orbital simulation of 2021 PH27 out to 100,000
  years. The orbit of 2021 PH27 is dynamically unstable over a few
  million years with numerous close encounters with Venus.  The
  coupling of the eccentricity and inclination overtime suggests 2021
  PH27 is experiencing a resonant interaction with the planets.
\label{fig:2021PH27stability2} }
\end{figure}

The orbit of 2021 PH27 is dynamically unstable within a few million
year time-scale (Figures~\ref{fig:2021PH27stability1}
and~\ref{fig:2021PH27stability2}). There is a 0.7\% chance that it
will collide with Venus in the next million years, and it will pass
within Venus' Hill Sphere in the next 950 to 1050 years as all of our
clones travel through Venus' Hill Sphere in that time.  2021 PH27's
very close perihelion to the Sun means 2021 PH27 experiences the
largest General Relativistic effects on any known object in our Solar
System, including Mercury.  We used the ReboundX code to determine the
General Relativistic effects on 2021 PH27 from the Sun (Tamayo et
al. 2020).  The orbital precession rate of 2021 PH27 is about 53
arcseconds per century, which is faster than Mercury's orbital
precession rate of about 43 arcseconds per century (Will 2018).  The
General Relativity effects are not important in 2021 PH27's orbital
evolution as planetary perturbations are much more significant. The
asteroid interacts strongly with Venus and is likely in some sort of
Kozai-Lidov oscillation from Jupiter and/or the inner planets as seen
by the coupling of the eccentricity and inclination of the evolution
of 2021 PH27's orbit in Figure~\ref{fig:2021PH27stability2} (Kozai
1962; Lidov 1962; de la Fuente Marcos \& de la Fuente Marcos
2021). 2021 PH27's orbit and Venus overlap enough that it is a
Potentially Hazardous Asteroid to Venus with a Venus Minimum Orbit
Intersection Distance (MOID) of only 0.015. If 2021 PH27 is or becomes
an active asteroid, it could create meteor showers in Venus'
atmosphere (Albino et al. 2022).

2021 PH27 can reach surface temperatures of around 500 degrees Celsius
during its closest approach to the Sun.  This means its surface has
been strongly thermally processed over time and its internal structure
likely has been heavily stressed from the intense and changing thermal
environment (Li \& Jewitt 2013; Lisse \& Steckloff 2022).

2021 PH27 was relatively bright at discovery, being some 19.2
magnitudes in the r-band. As the aphelion distance is only 0.79 au,
which is just beyond Venus' orbit, this object, though relatively
bright and around 1 km in size, would be very hard for most NEO
surveys to find as they generally do not observe near Venus' orbit or
interior to it.  The large inclination of about 32 degrees also means
2021 PH27 spends most of its time well away from the ecliptic.

Though 2021 PH27 most likely came from the main asteroid belt, it is
possible that it could have originated much closer to the Sun from a
possible stable reservoir of small objects in resonance with Venus or
even the hypothetical Vulcanoid population, which is a theoretical
stable area of small bodies interior to Mercury (Greenstreet et
al. 2012).

\subsection{2021 LJ4: Atira Type}

2021 LJ4 was found in twilight images taken on UT 2021 June 6 using
DECam (Sheppard and Tholen 2021).  The object was recovered again
using DECam on UT 2021 June 8 as well as on UT 2021 June 9 and 11
using the CFHT telescope on Mauna Kea.  Additional recoveries were
later made using DECam.  The object was faint at some 21.4 mags in the
r-band at discovery. The asteroid appears to be a typical Atira type
with a period of 0.55 years and an aphelion just interior to Earth's
orbit at 0.93 au, with it crossing the orbits of both Venus and
Mercury (Table 2). 2021 LJ4 was recovered in November 2021 as well as
May 2022, yielding a well determined orbit for the object.  Assuming a
moderate S-type asteroid albedo, the object is likely to be 300 to 400
meters in diameter. Though 2021 LJ4 has only a moderate inclination,
its Atira orbit and somewhat small size make it hard for most NEO
surveys using smaller telescopes to detect this asteroid, making it an
ideal discovery using DECam during twilight observing.

\subsection{2022 AP7: Apollo Type}

The object 2022 AP7 was discovered using DECam in twilight on UT 2022
January 13 (Sheppard 2022).  It was recovered using DECam a few nights
later on UT 2022 January 16 using the asteroid short-arc orbit and
ephemeris computation program (known as KNOBS) written by D. Tholen
(Tholen \& Whiteley 2000).  Recovery occurred again using DECam on the
nights of UT 2022 January 18, 21 and 23. After recovery a few weeks
later using the Las Cumbres telescopes and the University of Hawaii
88-inch telescope, 2022 AP7 was found in data from 2017 in both the
NEO Pan-STARRS and Catalina Sky Survey images, when the object was
last near opposition, though far from Earth and thus faint.  The orbit
is thus very well known based on a $\sim5$ year observation arc.  2022
AP7 is an Apollo type NEO which crosses the Earth's orbit with a
perihelion near 0.83 au and aphelion near Jupiter at 5.0 au (Table 2).

The Earth Minimum Orbit Intersection Distance (MOID) for 2022 AP7 is
only 0.0475 au, making it a Potentially Hazardous Asteroid (PHA) and
likely the largest PHA found since 2014 based on absolute magnitude.
2022 AP7 is likely to be in the top $5\%$ of the largest PHAs
known. 2022 AP7 was relatively faint at discovery being 20.8
magnitudes, but because it was relatively far from the Earth at about
1.9 au and distant from the Sun around 1.4 au, it is a fairly large
object, likely being well over 1 km in size assuming a moderate albedo
(1.0 to 2.3 km diameter for an albedo of 0.25 to 0.05 respectively).

Such a large object with an orbital period of only 5 years and only a
moderate inclination might be expected to have been found earlier by
one of the NEO surveys that cover most of the sky exterior to Earth's
orbit.  Thus either 2022 AP7 has an orbit that aliases Earth's orbit,
having it usually at a large distance from Earth when in the night sky
near opposition, or 2022 AP7 may be brightening as it comes to
perihelion from cometary effects.  2022 AP7 was found as it approached
perihelion, which is when cometary activity is expected to increase
significantly.  In our discovery and recovery images, no obvious coma
or tail was detected.  As the orbital period of 2022 AP7 is near
exactly 5 years, it does have an orbit that currently aliases with
Earth's, keeping it well away from Earth when near opposition for now,
meaning it would only be efficiently found in a twilight type survey
as it would be near the Sun and brightest only when Earth is more on
the other side of the Sun and its elongation very low.  Thus 2022 AP7
is a discovery that exemplifies how a relatively large telescope
observing towards the Sun during twilight can find large NEOs that
most of the current NEO surveys do not efficiently find.  Many of the
``missing'' yet-to-be-found $\sim 1$ km sized NEOs likely have orbits
that alias with Earth, making them distant and faint when in the night
sky at opposition like 2022 AP7.

\section{Discussion}

The main design of the survey was to find or put constraints on the
population and orbital properties of objects near to or interior to
Venus' orbit.  We did not find any relatively stable Venus co-orbital
or resonant objects, allowing us to put strong upper limits on the
size of such a population if it exists. Though not the main goal of
the survey, four Atira type asteroids were detected, which is $\sim
15\%$ of the known Atira population. This allows us to put significant
constraints on the size of the Atira population, which have orbits
completely interior to Earth's.  In addition, this survey is one of
the few that can put modest constraints on the Vatira population,
which have orbits completely interior to Venus' orbit.

To determine what percentage of each interior population's orbits we
may have surveyed, we performed several Monte-Carlo simulations.
These simulations consisted of generating 100,000 clones of orbits of
a particular population and then determining how many of these objects
our survey would have been expected to detect based on the fields
observed, limiting magnitude of each field and the efficiency of the
survey.  This gives a rough percentage of the entire population of
orbits.  We then use Poisson statistics to put limits or upper limits,
in the case of no detections, of each population simulated.

For each simulation we assumed asteroids of 6 spectral types (S-type
with 0.22 albedo, M with 0.17, E with 0.45, C with 0.06, P with 0.04,
and D with 0.05 (Mainzer et al. 2011, 2012; Shevchenko et al. 2016;
Morbidelli et al. 2020)) and 3 diameters ($D=0.5, 1.0$ and $1.5$ km)
using the same methods as detailed in Pokorny et al. (2020). See
Pokorny et al. (2020) for more details on our simulations, including
the phase function, apparent visual magnitude, and diameter
conversions we used in our simulations.  The asteroid type and hence
the albedo of the asteroid only matters for detecting objects less
than 1 km as all objects, no matter their type, would have similar
detection efficiencies near or above 1 km as our survey was deep
enough to make almost all 1 km or bigger inner asteroids efficiently
detected.

\subsection{Stable Venus Resonant Objects}

Pokorny \& Kuchner (2019) found that some low-eccentricity,
low-inclination Venus co-orbital objects could be in resonance and
stable for the age of the Solar System.  If this population exists, it
would be a very interesting population to find and study as these
objects could be the remnant of the inner solar system's formation
disk.  Pokorny et al. (2020) performed a limited survey with DECam of
the space near Venus' orbit to put modest constraints on this possible
population of stable Venus resonant co-orbitals.  Here we extend the
Pokorny et al. (2020) analysis using the new, much larger survey area
covered near Venus' orbit with DECam. For this population we followed
100,000 clones of low-eccentricity, low-inclination, Venus co-orbital
asteroids that were shown to be stable for the age of the solar system
in Pokorny and Kuchner (2019).

Based on our simulations, this DECam twilight survey should have
observed some 21 to 23 percent of the entire population of stable Venus
co-orbitals larger than 1 km, which depends on the spectral type and
thus albedo of the objects (Table 3). Since we found no relatively
stable Venus resonant objects and there are none known, this
population is not likely to be very large.  We calculate the upper
limit on the population size of 1 km or larger stable Venus
co-orbitals to be only about $4^{+6}_{-3}$, depending on the asteroid
type (Table 3).  Similar constraints are found for even smaller 0.5 km
sized higher albedo asteroids types like the S, M and E-type, as they
would be relatively easier to find because of their high reflectance.
For the darker asteroid types, like the C, P and D-types, the smaller
sized asteroids of 0.5 km are basically unconstrained as most would be
too faint for our survey to efficiently detect (Table 3).

\subsection{Unstable Venus Near Co-orbitals}

There are five well observed asteroids that currently have orbital
periods similar to Venus: (322756) 2001 CK32, (524522) 2002 VE68, 2012
XE133, 2013 ND15 and 2015 WZ12. All of these, as well as the newer
objects found since 2020 are much smaller than 1 km and are on
dynamically unstable orbits on the order of a few million years, as
discussed in the introduction.  We generated 100,000 clones using the
above five object's semi-major axes, eccentricities and inclinations
with randomly selecting the remaining orbital elements. Though we did
not observe any of these known objects, through this simulation, we
find our survey would have detected about 5 to 10 percent of the
objects similar to these known dynamically unstable Venus near
co-orbitals and larger than 1 km in size, depending on their albedos
and spectral types (Table 4).  From this simulation there could be a
few of these near unstable Venus co-orbitals to discover around the 1
km size regime.


\subsection{Atira/Apohele Asteroid Population}

In addition to the two new Atira asteroids discovered in this survey,
2021 LJ4 and 2021 PH27, we also detected serendipitously the
previously known Atira asteroids 2019 AQ3 and 2021 BS1. We used the
semi-major axis, eccentricity and inclination of the 25 known Atira
objects with well-determined orbits as of March 2022 and randomly
selected the other orbital elements to generate 100,000 clones.  Table
5 shows the percentage of the population of Atira objects our survey should
have detected, which amounts to 5 to 8 percent of the total population
larger than 1 km, depending on the albedos of the objects.

Our simulations show there are about 50 to 75 $\pm 35$ Atiras that are
about 1 km or larger (Table 5), suggesting less than half of the
largest objects in the Atira population have been found to
date. Assuming there are about 1000 NEOs of 1 km or larger, this would
make the Atira population about 5\% of the NEO population.  This
result generally agrees with estimates of the Atira population size
from some earlier NEO models (Bottke et al. 2002; Greenstreet et al
2012), but is somewhat higher than the currently modelled $\sim 1\%$
for the Atira population fraction of NEOs (Granvik et al. 2018).

These results are based on low number statistics and mostly assumed
albedos, so if the population size of 1 km Atira objects is on the low
end of the uncertainty, than there are likely only a few large Atiras
left to find. The larger undiscovered Atira asteroids that remain
probably have high inclinations and/or smaller than average semi-major
axes and/or eccentricities, which would keep them mostly well interior
to the Earth's orbit and away from the darker skies where the main NEO
surveys operate most efficiently.  This is true for the 1 km sized
Atira 2021 PH27, which we found with a high inclination and very low
semi-major axis (Table 2).

\subsection{Vatira Asteroid Population}

Vatira asteroids have orbits completely interior to Venus' orbit.
Since the Vatira population's orbital distribution is still highly
uncertain, we simulated two different Vatira populations to understand
how sensitive our survey would be to finding Vatira asteroids.  First,
we used the semi-major axis, eccentricity and inclination of the only
known Vatira asteroid (594913) 'Ayl\'{o}'chaxnim (2020 AV2) and
generated 100,000 clones by randomly selecting the remaining orbital
elements. In this simulation, we would have found about 5 percent of
the $\sim 1$ km or larger Vatira asteroids with similar orbits as 2020
AV2 (Table 6). This puts only minimal constraints on a Vatira
population with orbits like 2020 AV2, as there could still be tens of
objects like 2020 AV2 to be found since they spend most of their time
at smaller elongations than many of our survey fields.

In a second simulation, we made a 100,000 asteroid hypothetical Vatira
population by creating a somewhat circular and moderately inclined
population through randomly selecting orbital elements with $0.50 < a
< 0.60$ au, $0.0 < e < 0.1$, $0<i<15$ degrees and randomly selecting
the other remaining orbital elements between 0 and 360 degrees. As
many of the objects in this hypothetical Vatira population would not
have elongations as great as 2020 AV2, it would be much harder to
detect these objects in our fields. Thus our survey would only find at
most about 1 percent of objects in this hypothetical Vatira population
(Table 7).

\section{Conclusions and Summary}

The twilight survey using DECam on the CTIO Blanco 4 meter telescope
is one of the largest area and sensitive searches ever performed for
objects interior to Earth's and near Venus' orbit. Though no objects
with orbits similar to or interior to Venus' orbit were found, the
survey did find three relatively large NEOs, including two Atira and
one Apollo orbital type. The new discovery 2021 PH27 is about 1 km in
size and has the smallest semi-major axis and thus shortest orbital
period around the Sun of any known asteroid.  2021 PH27 has strong
interactions with Venus, with it likely passing through Venus' Hill
Sphere in 950 to 1050 years from now, making it a potentially
hazardous asteroid to Venus. The eccentric orbit of 2021 PH27 means it
crosses both Mercury's and Venus' orbit and its very low perihelion of
only 0.13 au creates about a 53 arcsecond per century precession in
its orbit from General Relativistic effects, which are the largest
known in our Solar System. The newly discovered 2022 AP7 is an Apollo
type NEO that is probably the largest Potentially Hazardous NEO to
Earth found in several years, being some 1.5 km in size. There are
likely several more 1 km sized Atira type asteroids left to find,
which probably have low semi-major axes and high inclinations, like
2021 PH27, making them hard to find for most asteroid surveys.  The
DECam twilight survey is covering sky geometries and areas that most
other surveys do not cover to depths not usually obtained, filling an
important niche in the survey for the last few remaining relatively
large unknown NEOs.  Interestingly, the twilight survey has discovered
more larger asteroids ($\gtrsim 1$ km) than smaller ones even though
the survey is sensitive to smaller asteroids.  This might suggest the
smaller asteroids are dynamically less stable and/or more susceptible
to break-up from the extreme thermal and gravitational environment
near the Sun, though additonal discoveries of asteriods with orbits
near the Sun must be made to statistically determine if the smaller
asteroids are under-abundant since in general they are also harder to
detect.


\section*{Acknowledgments}

Observations were obtained at Cerro Tololo Inter-American Observatory,
National Optical Astronomy Observatory, which are operated by the
Association of Universities for Research in Astronomy, under contract
with the National Science Foundation.  This project used data obtained
with the Dark Energy Camera (DECam), which was constructed by the Dark
Energy Survey (DES) collaborating institutions: Argonne National Lab,
University of California Santa Cruz, University of Cambridge, Centro
de Investigaciones Energeticas, Medioambientales y
Tecnologicas-Madrid, University of Chicago, University College London,
DES-Brazil consortium, University of Edinburgh, ETH-Zurich, University
of Illinois at Urbana-Champaign, Institut de Ciencies de l'Espai,
Institut de Fisica d'Altes Energies, Lawrence Berkeley National Lab,
Ludwig-Maximilians Universitat, University of Michigan, National
Optical Astronomy Observatory, University of Nottingham, Ohio State
University, University of Pennsylvania, University of Portsmouth, SLAC
National Lab, Stanford University, University of Sussex, and Texas
A\&M University. Funding for DES, including DECam, has been provided
by the U.S. Department of Energy, National Science Foundation,
Ministry of Education and Science (Spain), Science and Technology
Facilities Council (UK), Higher Education Funding Council (England),
National Center for Supercomputing Applications, Kavli Institute for
Cosmological Physics, Financiadora de Estudos e Projetos, Fundação
Carlos Chagas Filho de Amparo a Pesquisa, Conselho Nacional de
Desenvolvimento Científico e Tecnológico and the Ministério da Ciência
e Tecnologia (Brazil), the German Research Foundation-sponsored
cluster of excellence "Origin and Structure of the Universe" and the
DES collaborating institutions. DT was supported by NASA grant
80NSSC21K0807. PP was supported by NASA ISFM EIMM award, the NASA
Cooperative Agreement 80GSFC21M0002 and NASA Solar System Workings
award 80NSSC21K0153. TSR acknowledges funding from the NEO-MAPP
project (H2020-EU-2-1-6/870377).  CM was partially supported by the
international Gemini Observatory, a program of NSF’s NOIRLab, which is
managed by the Association of Universities for Research in Astronomy
(AURA) under a cooperative agreement with the National Science
Foundation, on behalf of the Gemini partnership of Argentina, Brazil,
Canada, Chile, the Republic of Korea, and the United States of
America.  This work was (partially) funded by the Spanish
MICIN/AEI/10.13039/501100011033 and by “ERDF A way of making Europe”
by the “European Union” through grant RTI2018-095076-B-C21, and the
Institute of Cosmos Sciences University of Barcelona (ICCUB, Unidad de
Excelencia ‘Mar\'ia de Maeztu’) through grant CEX2019-000918-M.  This
work makes use of observations from the Las Cumbres Observatory global
telescope network. This paper includes data gathered with the 6.5
meter Magellan Telescopes located at Las Campanas Observatory, Chile.

\clearpage

\newpage

\clearpage

\begin{center}
\startlongtable
\begin{deluxetable}{lcccc}
\tablenum{1}
\tablewidth{7 in}
\tablecaption{Observed Twilight DECam Fields}
\tablecolumns{5}
\tablehead{
\colhead{RA(J2000)} & \colhead{Dec(J2000)} & \colhead{UT Date} & \colhead{Limiting} & \colhead{Seeing} \\ \colhead{hh:mm:ss} & \colhead{dd:mm:ss} & \colhead{YYYY/MM/DD/hh:mm} & \colhead{(mag)} & \colhead{(arcsec)}}  
\startdata
14:45:00  &  -17:30:00  &   2019/09/23/23:30  & 21.0 & 1.3  \\
14:45:00  &  -19:30:00  &   2019/09/23/23:31  & 21.0 & 1.3  \\
14:49:00  &  -17:30:00  &   2019/09/23/23:32  & 21.0 & 1.3  \\
14:49:00  &  -19:30:00  &   2019/09/23/23:33  & 21.0 & 1.3  \\
14:49:00  &  -15:30:00  &   2019/09/23/23:34  & 21.0 & 1.3  \\
14:51:00  &  -16:15:00  &   2019/09/25/23:30  & 21.0 & 1.3  \\
14:51:00  &  -18:15:00  &   2019/09/25/23:31  & 21.0 & 1.3  \\
14:51:00  &  -20:15:00  &   2019/09/25/23:32  & 21.0 & 1.3  \\
14:59:00  &  -16:15:00  &   2019/09/25/23:33  & 21.0 & 1.3  \\
14:59:00  &  -18:15:00  &   2019/09/25/23:34  & 21.0 & 1.3  \\
14:49:00  &  -17:30:00  &   2019/09/26/23:30  & 21.0 & 1.3  \\
14:49:00  &  -19:30:00  &   2019/09/26/23:31  & 21.0 & 1.3  \\
14:49:00  &  -15:30:00  &   2019/09/26/23:32  & 21.0 & 1.3  \\
14:58:00  &  -15:30:00  &   2019/09/26/23:33  & 21.0 & 1.3  \\
14:58:00  &  -17:30:00  &   2019/09/26/23:34  & 21.0 & 1.3  \\
14:47:00  &  -17:30:00  &   2019/09/27/23:30  & 21.0 & 1.3  \\
14:47:00  &  -19:30:00  &   2019/09/27/23:31  & 21.0 & 1.3  \\
14:56:00  &  -17:30:00  &   2019/09/27/23:32  & 21.0 & 1.3  \\
14:56:00  &  -19:30:00  &   2019/09/27/23:33  & 21.0 & 1.3  \\
14:46:00  &  -17:30:00  &   2019/09/28/23:30  & 21.0 & 1.3  \\
14:46:00  &  -19:30:00  &   2019/09/28/23:31  & 21.0 & 1.3  \\
14:46:00  &  -21:15:00  &   2019/09/28/23:32  & 21.0 & 1.3  \\
14:59:00  &  -21:15:00  &   2019/09/28/23:33  & 21.0 & 1.3  \\
14:59:00  &  -19:30:00  &   2019/09/28/23:34  & 21.0 & 1.3  \\
08:15:00  &  +14:00:00  &   2021/06/06/22:55  & 21.5 & 1.0  \\
08:15:00  &  +16:00:00  &   2021/06/06/22:56  & 21.6 & 1.0  \\
08:15:00  &  +12:00:00  &   2021/06/06/22:57  & 21.7 & 1.0  \\
08:22:00  &  +16:00:00  &   2021/06/06/22:58  & 21.7 & 1.0  \\
08:22:00  &  +14:00:00  &   2021/06/06/22:59  & 21.7 & 1.0  \\
08:22:00  &  +12:00:00  &   2021/06/06/23:00  & 21.7 & 1.0  \\
08:22:00  &  +17:00:00  &   2021/06/07/22:55  & 21.4 & 1.0  \\
08:15:00  &  +10:00:00  &   2021/06/07/22:56  & 21.5 & 1.0  \\
08:10:00  &  +08:00:00  &   2021/06/07/22:57  & 21.6 & 1.0  \\
08:10:00  &  +10:00:00  &   2021/06/07/22:58  & 21.6 & 1.0  \\
08:10:00  &  +12:00:00  &   2021/06/07/22:59  & 21.7 & 1.0  \\
08:30:00  &  +16:00:00  &   2021/06/07/23:00  & 21.7 & 1.0  \\
08:14:00  &  +16:00:00  &   2021/06/09/22:55  & 20.7 & 1.5  \\
08:14:00  &  +14:00:00  &   2021/06/09/22:56  & 20.7 & 1.5  \\
08:20:00  &  +17:50:00  &   2021/06/10/22:55  & 21.0 & 1.4  \\
08:28:00  &  +18:10:00  &   2021/06/10/22:56  & 21.0 & 1.4  \\
08:21:00  &  +10:34:00  &   2021/06/10/22:57  & 21.0 & 1.4  \\
08:02:00  &  +10:00:00  &   2021/06/12/22:55  & 21.4 & 1.3  \\
07:57:00  &  +08:00:00  &   2021/06/12/22:56  & 21.5 & 1.3  \\
07:52:00  &  +06:04:00  &   2021/06/12/22:57  & 21.5 & 1.3  \\
07:48:00  &  +04:00:00  &   2021/06/12/22:58  & 21.5 & 1.3  \\
08:25:53  &  +10:32:00  &   2021/06/12/22:59  & 21.7 & 1.3  \\
07:51:00  &  +02:00:00  &   2021/06/13/22:55  & 21.6 & 1.1  \\
07:46:00  &  +00:01:00  &   2021/06/13/22:56  & 21.7 & 1.1  \\
08:50:00  &  +12:34:00  &   2021/06/13/22:57  & 21.8 & 1.1  \\
08:50:00  &  +14:30:00  &   2021/06/13/22:58  & 21.9 & 1.1  \\
08:51:55  &  +19:50:00  &   2021/06/13/22:59  & 21.9 & 1.1  \\
05:41:00  &  +06:00:00  &   2021/07/25/10:25  & 20.7 & 1.9  \\
05:35:00  &  +08:00:00  &   2021/07/25/10:26  & 20.7 & 1.9  \\
05:29:00  &  +10:00:00  &   2021/07/25/10:27  & 20.7 & 1.9  \\
05:21:00  &  +12:00:00  &   2021/07/25/10:28  & 20.7 & 1.9  \\
10:40:00  &  -01:00:00  &   2021/07/25/23:07  & 21.3 & 1.7  \\
10:35:00  &  -03:00:00  &   2021/07/25/23:08  & 21.3 & 1.7  \\
10:29:00  &  -05:00:00  &   2021/07/25/23:09  & 21.3 & 1.7  \\
10:25:00  &  -07:00:00  &   2021/07/25/23:10  & 21.3 & 1.7  \\
05:27:00  &  +13:00:00  &   2021/07/26/10:24  & 20.8 & 1.8  \\
05:19:30  &  +15:00:00  &   2021/07/26/10:25  & 20.8 & 1.8  \\
05:11:00  &  +17:00:00  &   2021/07/26/10:26  & 20.8 & 1.8  \\
05:04:00  &  +18:30:00  &   2021/07/26/10:27  & 20.8 & 1.8  \\
11:07:30  &  +11:00:00  &   2021/07/26/23:05  & 21.1 & 1.6  \\
11:07:30  &  +09:00:00  &   2021/07/26/23:06  & 21.4 & 1.6  \\
11:02:00  &  +07:00:00  &   2021/07/26/23:07  & 21.4 & 1.6  \\
10:57:00  &  +05:00:00  &   2021/07/26/23:08  & 21.4 & 1.6  \\
10:51:00  &  +03:00:00  &   2021/07/26/23:09  & 21.4 & 1.6  \\
10:47:00  &  +01:00:00  &   2021/07/26/23:10  & 21.4 & 1.6  \\
05:57:00  &  +05:00:00  &   2021/07/27/10:25  & 20.3 & 1.9  \\
06:01:30  &  +03:00:00  &   2021/07/27/10:26  & 20.3 & 1.9  \\
06:09:30  &  +01:00:00  &   2021/07/27/10:27  & 20.3 & 1.9  \\
11:17:00  &  +12:30:00  &   2021/07/27/23:06  & 20.1 & 2.5  \\
10:39:30  &  -01:00:00  &   2021/07/27/23:07  & 20.5 & 2.5  \\
10:36:30  &  -03:00:00  &   2021/07/27/23:08  & 20.5 & 2.5  \\
11:06:30  &  +04:00:00  &   2021/07/27/23:09  & 20.5 & 2.5  \\
11:12:30  &  +06:00:00  &   2021/07/27/23:10  & 20.5 & 2.5  \\
06:29:00  &  -05:00:00  &   2021/07/28/10:25  & 21.1 & 1.6  \\
06:25:00  &  -03:00:00  &   2021/07/28/10:26  & 21.1 & 1.6  \\
06:19:30  &  -01:00:00  &   2021/07/28/10:27  & 21.1 & 1.6  \\
07:11:00  &  -05:00:00  &   2021/08/10/10:13  & 20.9 & 1.3  \\
07:21:00  &  -06:10:00  &   2021/08/11/10:20  & 21.0 & 1.4  \\
07:15:00  &  -03:00:00  &   2021/08/11/10:21  & 21.0 & 1.4  \\
07:10:00  &  -01:00:00  &   2021/08/11/10:22  & 21.0 & 1.4  \\
07:35:00  &  -08:00:00  &   2021/08/13/10:18  & 20.6 & 1.4  \\
07:40:00  &  -10:00:00  &   2021/08/13/10:19  & 20.4 & 1.4  \\
11:45:00  &  -03:00:00  &   2021/08/13/23:08  & 20.5 & 1.7  \\
11:51:50  &  -02:46:10  &   2021/08/14/23:09  & 20.4 & 1.6  \\
11:57:30  &  -01:00:00  &   2021/08/14/23:10  & 20.4 & 1.6  \\
12:02:30  &  +01:00:00  &   2021/08/14/23:11  & 20.4 & 1.6  \\
12:37:00  &  -10:00:00  &   2021/08/28/23:21  & 21.8 & 1.4  \\
12:41:00  &  -08:00:00  &   2021/08/28/23:22  & 21.8 & 1.4  \\
12:46:00  &  -06:00:00  &   2021/08/28/23:23  & 21.8 & 1.4  \\
12:54:20  &  -04:00:00  &   2021/08/28/23:24  & 21.8 & 1.4  \\
12:59:30  &  -02:00:00  &   2021/08/28/23:25  & 21.8 & 1.4  \\
13:05:10  &  -00:10:00  &   2021/08/28/23:26  & 21.8 & 1.4  \\
12:43:00  &  -12:00:00  &   2021/08/29/23:22  & 21.1 & 1.9  \\
12:38:38  &  -09:35:55  &   2021/08/29/23:23  & 21.1 & 1.9  \\
13:08:53  &  -02:27:30  &   2021/08/29/23:24  & 21.1 & 1.9  \\
13:17:00  &  +01:40:00  &   2021/08/30/23:25  & 20.2 & 2.2  \\
13:22:00  &  +03:25:00  &   2021/08/30/23:26  & 20.2 & 2.2  \\
12:50:00  &  -16:00:00  &   2021/08/31/23:23  & 19.9 & 1.6  \\
12:45:00  &  -18:00:00  &   2021/08/31/23:24  & 19.9 & 1.6  \\
14:12:00  &  -06:00:00  &   2021/09/16/23:29  & 20.7 & 1.5  \\
14:12:30  &  -08:00:00  &   2021/09/16/23:30  & 21.0 & 1.5  \\
14:08:00  &  -10:00:00  &   2021/09/16/23:31  & 21.0 & 1.5  \\
14:04:00  &  -12:00:00  &   2021/09/16/23:32  & 21.0 & 1.5  \\
14:12:00  &  -14:00:00  &   2021/09/19/23:31  & 20.0 & 2.6  \\
14:17:00  &  -18:20:00  &   2021/09/22/23:31  & 21.2 & 1.3  \\
14:20:00  &  -16:40:00  &   2021/09/22/23:32  & 21.2 & 1.3  \\
14:24:00  &  -15:00:00  &   2021/09/22/23:33  & 21.2 & 1.3  \\
14:28:00  &  -13:20:00  &   2021/09/22/23:34  & 21.2 & 1.3  \\
15:06:00  &  -07:30:00  &   2021/09/27/23:31  & 21.6 & 1.3  \\
15:02:00  &  -09:20:00  &   2021/09/27/23:32  & 21.6 & 1.3  \\
14:57:00  &  -11:10:00  &   2021/09/27/23:33  & 21.6 & 1.3  \\
15:45:00  &  -27:00:00  &   2021/10/16/23:50  & 21.0 & 1.4  \\
16:02:00  &  -21:00:00  &   2021/10/16/23:52  & 21.3 & 1.4  \\
15:58:00  &  -23:00:00  &   2021/10/16/23:53  & 21.3 & 1.4  \\
15:54:00  &  -25:00:00  &   2021/10/16/23:54  & 21.3 & 1.4  \\
16:26:00  &  -13:00:00  &   2021/10/19/23:50  & 21.0 & 1.2  \\
16:27:00  &  -15:00:00  &   2021/10/19/23:51  & 21.2 & 1.2  \\
16:24:00  &  -17:00:00  &   2021/10/19/23:52  & 21.2 & 1.2  \\
16:20:00  &  -19:00:00  &   2021/10/19/23:53  & 21.2 & 1.2  \\
10:29:00  &  -05:00:00  &   2021/10/20/08:52  & 20.7 & 1.9  \\
10:33:00  &  -07:00:00  &   2021/10/20/08:53  & 20.7 & 1.9  \\
10:38:00  &  -09:00:00  &   2021/10/20/08:54  & 20.7 & 1.9  \\
10:45:00  &  -11:00:00  &   2021/10/21/08:51  & 21.0 & 1.7  \\
10:49:00  &  -13:00:00  &   2021/10/21/08:52  & 21.0 & 1.7  \\
10:53:00  &  -15:00:00  &   2021/10/21/08:53  & 21.0 & 1.7  \\
16:09:00  &  -34:00:00  &   2021/10/22/23:50  & 21.4 & 1.2  \\
16:16:00  &  -32:00:00  &   2021/10/22/23:51  & 21.7 & 1.2  \\
16:16:00  &  -30:00:00  &   2021/10/22/23:52  & 21.7 & 1.2  \\
16:19:00  &  -28:20:00  &   2021/10/22/23:53  & 21.7 & 1.2  \\
11:02:00  &  -17:00:00  &   2021/10/23/08:48  & 20.2 & 1.8  \\
11:06:00  &  -19:00:00  &   2021/10/23/08:49  & 20.2 & 1.8  \\
11:10:00  &  -21:00:00  &   2021/10/23/08:50  & 20.2 & 1.8  \\
10:41:00  &  -04:00:00  &   2021/10/25/08:46  & 21.6 & 1.1  \\
10:35:00  &  -02:00:00  &   2021/10/25/08:47  & 21.6 & 1.1  \\
10:30:00  &  -00:10:00  &   2021/10/25/08:48  & 21.6 & 1.1  \\
11:19:00  &  -07:00:00  &   2021/11/04/08:39  & 20.6 & 1.9  \\
11:24:30  &  -09:00:00  &   2021/11/04/08:40  & 20.6 & 1.9  \\
11:29:00  &  -11:00:00  &   2021/11/04/08:41  & 20.6 & 1.9  \\
11:39:00  &  -13:00:00  &   2021/11/04/08:46  & 20.4 & 1.9  \\
11:51:30  &  -15:00:00  &   2021/11/07/08:41  & 20.0 & 1.7  \\
11:56:00  &  -17:00:00  &   2021/11/07/08:42  & 20.0 & 1.7  \\
12:00:00  &  -19:00:00  &   2021/11/07/08:43  & 20.0 & 1.7  \\
20:39:00  &  -26:00:00  &   2021/12/16/00:45  & 21.3 & 1.3  \\
20:57:00  &  -20:00:00  &   2021/12/16/00:46  & 21.5 & 1.3  \\
20:53:30  &  -22:00:00  &   2021/12/16/00:47  & 21.5 & 1.3  \\
20:50:00  &  -24:00:00  &   2021/12/16/00:48  & 21.5 & 1.3  \\
14:14:00  &  -20:00:00  &   2021/12/16/08:20  & 21.7 & 1.1  \\
14:18:30  &  -22:00:00  &   2021/12/16/08:21  & 21.7 & 1.1  \\
14:23:00  &  -24:00:00  &   2021/12/16/08:22  & 21.7 & 1.1  \\
14:32:00  &  -26:00:00  &   2021/12/16/08:27  & 21.4 & 1.1  \\
20:45:20  &  -32:00:00  &   2021/12/19/00:46  & 20.8 & 1.8  \\
20:52:00  &  -30:30:00  &   2021/12/19/00:48  & 21.0 & 1.8  \\
20:53:30  &  -29:00:00  &   2021/12/19/00:49  & 21.0 & 1.8  \\
20:56:30  &  -27:30:00  &   2021/12/19/00:50  & 21.0 & 1.8  \\
14:42:00  &  -27:30:00  &   2021/12/19/08:22  & 21.0 & 1.8  \\
14:44:30  &  -29:00:00  &   2021/12/19/08:23  & 21.0 & 1.8  \\
14:46:00  &  -30:30:00  &   2021/12/19/08:24  & 21.0 & 1.8  \\
14:52:00  &  -32:00:00  &   2021/12/19/08:27  & 20.8 & 1.8  \\
21:24:20  &  -15:30:00  &   2021/12/22/00:47  & 21.1 & 1.5  \\
21:27:00  &  -17:00:00  &   2021/12/22/00:48  & 21.3 & 1.5  \\
21:25:00  &  -18:30:00  &   2021/12/22/00:49  & 21.3 & 1.5  \\
21:22:30  &  -20:00:00  &   2021/12/22/00:50  & 21.3 & 1.5  \\
21:20:00  &  -21:30:00  &   2021/12/22/00:51  & 21.3 & 1.5  \\
14:32:00  &  -16:00:00  &   2021/12/22/08:25  & 21.0 & 1.8  \\
14:36:00  &  -17:30:00  &   2021/12/22/08:26  & 21.0 & 1.8  \\
14:40:00  &  -19:00:00  &   2021/12/22/08:27  & 21.0 & 1.8  \\
14:47:00  &  -20:30:00  &   2021/12/22/08:28  & 20.8 & 1.8  \\
14:49:00  &  -22:00:00  &   2021/12/23/08:25  & 21.4 & 1.3  \\
14:52:00  &  -23:30:00  &   2021/12/23/08:26  & 21.4 & 1.3  \\
14:55:00  &  -25:00:00  &   2021/12/23/08:27  & 21.4 & 1.3  \\
15:00:00  &  -26:30:00  &   2021/12/23/08:28  & 21.2 & 1.3  \\
21:08:00  &  -29:00:00  &   2021/12/24/00:47  & 21.3 & 1.2  \\
21:17:00  &  -27:30:00  &   2021/12/24/00:48  & 21.6 & 1.2  \\
21:20:00  &  -26:00:00  &   2021/12/24/00:49  & 21.6 & 1.2  \\
21:23:00  &  -24:30:00  &   2021/12/24/00:50  & 21.6 & 1.2  \\
21:26:00  &  -23:00:00  &   2021/12/24/00:51  & 21.6 & 1.2  \\
15:04:00  &  -28:00:00  &   2021/12/24/08:25  & 21.4 & 1.1  \\
15:07:00  &  -29:30:00  &   2021/12/24/08:26  & 21.4 & 1.1  \\
15:08:00  &  -31:15:00  &   2021/12/24/08:27  & 21.4 & 1.1  \\
15:13:00  &  -33:00:00  &   2021/12/24/08:31  & 21.1 & 1.1  \\
15:26:00  &  -34:00:00  &   2021/12/28/08:28  & 21.6 & 1.2  \\
15:26:30  &  -35:30:00  &   2021/12/28/08:29  & 21.6 & 1.2  \\
15:27:00  &  -37:00:00  &   2021/12/28/08:30  & 21.6 & 1.2  \\
15:32:00  &  -38:30:00  &   2021/12/28/08:34  & 21.3 & 1.2  \\
15:40:00  &  -40:00:00  &   2021/12/31/08:30  & 21.8 & 1.1  \\
15:40:30  &  -41:40:00  &   2021/12/31/08:31  & 21.8 & 1.1  \\
15:41:30  &  -43:30:00  &   2021/12/31/08:32  & 21.8 & 1.1  \\
15:47:00  &  -45:00:00  &   2021/12/31/08:35  & 21.5 & 1.1  \\
22:07:00  &  -20:00:00  &   2022/01/03/00:52  & 21.1 & 1.2  \\
15:37:00  &  -20:00:00  &   2022/01/03/08:34  & 22.0 & 1.1  \\
15:41:00  &  -21:30:00  &   2022/01/03/08:35  & 22.0 & 1.1  \\
15:45:00  &  -23:00:00  &   2022/01/03/08:36  & 21.9 & 1.1  \\
15:51:00  &  -24:30:00  &   2022/01/03/08:41  & 21.6 & 1.1  \\
16:02:00  &  -26:00:00  &   2022/01/06/08:34  & 22.0 & 0.9  \\
16:05:30  &  -27:30:00  &   2022/01/06/08:35  & 22.0 & 0.9  \\
16:08:30  &  -29:00:00  &   2022/01/06/08:36  & 21.9 & 0.9  \\
16:14:00  &  -30:30:00  &   2022/01/06/08:40  & 21.7 & 1.0  \\
16:15:00  &  -32:00:00  &   2022/01/06/08:41  & 21.6 & 1.0  \\
16:15:30  &  -25:00:00  &   2022/01/09/08:37  & 21.2 & 1.9  \\
16:13:30  &  -23:30:00  &   2022/01/09/08:38  & 20.8 & 2.2  \\
16:11:00  &  -22:00:00  &   2022/01/09/08:39  & 20.9 & 2.1  \\
16:12:00  &  -20:30:00  &   2022/01/09/08:44  & 20.4 & 2.0  \\
16:10:00  &  -19:00:00  &   2022/01/09/08:45  & 20.5 & 1.8  \\
22:43:00  &  -24:30:00  &   2022/01/13/00:58  & 21.3 & 1.2  \\
16:33:30  &  -24:00:00  &   2022/01/13/08:47  & 21.5 & 0.9  \\
16:31:30  &  -22:30:00  &   2022/01/13/08:48  & 21.6 & 0.9  \\
16:30:00  &  -21:00:00  &   2022/01/13/08:49  & 21.5 & 0.9  \\
16:29:00  &  -19:00:00  &   2022/01/13/08:53  & 21.3 & 0.9  \\
16:28:00  &  -17:30:00  &   2022/01/13/08:54  & 21.2 & 0.9  \\
16:43:16  &  -17:32:15  &   2022/01/18/08:49  & 21.1 & 1.4  \\
23:20:00  &  -20:00:00  &   2022/01/21/00:49  & 21.3 & 1.3  \\
23:23:30  &  -18:30:00  &   2022/01/21/00:50  & 21.4 & 1.3  \\
23:27:00  &  -17:00:00  &   2022/01/21/00:51  & 21.5 & 1.3  \\
23:30:30  &  -15:30:00  &   2022/01/21/00:52  & 21.5 & 1.3  \\
16:55:15  &  -17:34:30  &   2022/01/21/08:50  & 21.5 & 1.2  \\
17:03:00  &  -19:00:00  &   2022/01/21/08:51  & 21.5 & 1.2  \\
17:06:00  &  -20:30:00  &   2022/01/21/08:52  & 21.4 & 1.2  \\
17:12:00  &  -21:30:00  &   2022/01/21/08:57  & 21.2 & 1.2  \\
17:15:00  &  -23:00:00  &   2022/01/21/08:58  & 21.1 & 1.2  \\
17:02:00  &  -16:00:00  &   2022/01/22/08:52  & 21.2 & 1.5  \\
16:59:30  &  -14:30:00  &   2022/01/22/08:53  & 21.2 & 1.5  \\
16:57:30  &  -13:00:00  &   2022/01/22/08:54  & 21.2 & 1.5  \\
17:03:40  &  -17:34:30  &   2022/01/23/08:55  & 21.8 & 1.1  \\
16:57:00  &  -11:00:00  &   2022/01/23/08:57  & 21.8 & 1.1  \\
17:14:30  &  -17:00:00  &   2022/01/24/08:54  & 21.4 & 1.3  \\
16:59:50  &  -09:40:00  &   2022/01/24/08:55  & 21.4 & 1.3  \\
16:59:00  &  -08:00:00  &   2022/01/24/08:58  & 21.2 & 1.3  \\
\enddata
\tablenotetext{}{}
\end{deluxetable}
\end{center}

\clearpage

\newpage

\clearpage

%
%
%
%

\begin{center}
\begin{deluxetable}{lcccccc}
\tablenum{2}
\tablewidth{6.5 in}
\tablecaption{New Asteroid Discoveries}
\tablecolumns{7}
\tablehead{
\colhead{Name} & \colhead{a} & \colhead{e} & \colhead{i} & \colhead{H} &  \colhead{Diam} & \colhead{Type} \\ \colhead{} & \colhead{(AU)} & \colhead{} & \colhead{(deg)} & \colhead{(mag)} & \colhead{(km)} & \colhead{}}  
\startdata
2021 LJ4        &  0.676   & 0.382  & 9.827  & 20.1 & $0.3-0.6$  & Atira \\
2021 PH27       &  0.4617  & 0.712  & 31.927 & 17.7 & $0.9-1.7$  & Atira \\
2022 AP7        &  2.924   & 0.715  & 13.835 & 17.1 & $1.1-2.3$  & Apollo \\
\enddata
\tablenotetext{}{Orbital elements are from the Minor Planet Center as of June 2022, with the end number being the significant digit of the uncertainty. Though the albedos are unknown, they are likley between 5 to 20 percent as many NEOs are in this range (Pravec et al. 2012).  Using the standard 14 percent albedo assumption for unknown albedo NEOs as used by the NASA Center for Near Earth Object Studies, this means an absolute magnitude brighter than 17.75 mag is likely larger than 1 km in size. The diameter range shown above is assuming 20 percent and 5 percent albedos, respectively.}
\end{deluxetable}
\end{center}

%
%
%
%

\begin{center}
\begin{deluxetable}{lccc}
\tablenum{3}
\tablewidth{6 in}
\tablecaption{Stable Venus Co-orbital Completeness Percentage and Population}
\tablecolumns{4}
\tablehead{
\colhead{Type} & \colhead{D=0.5} & \colhead{D=1.0} & \colhead{D=1.5}  \\ \colhead{} & \colhead{(km)} & \colhead{(km)} & \colhead{(km)}}  
\startdata
S-type  &  22\%($4^{+6}_{-3}$)  &  23\%($4^{+6}_{-3}$)  &  23\%($4^{+6}_{-3}$)    \\
M-Type  &  21\%($4^{+7}_{-3}$)  &  23\%($4^{+6}_{-3}$)  &  23\%($4^{+6}_{-3}$)   \\
E-Type  &  23\%($4^{+6}_{-3}$)  &  23\%($4^{+6}_{-3}$)  &  23\%($4^{+6}_{-3}$)     \\
C-Type  &  0.8\%($100^{+170}_{-80}$)   &  21\%($4^{+7}_{-3}$)  &  23\%($4^{+6}_{-3}$)     \\
P-Type  &  0.2\%($540^{+900}_{-400}$)   &  20\%($4^{+8}_{-3}$)  &  23\%($4^{+6}_{-3}$)     \\
D-Type  &  0.4\%($210^{+350}_{-150}$)   &  21\%($4^{+7}_{-3}$)  &  23\%($4^{+6}_{-3}$)     \\
\enddata
\tablenotetext{}{Survey completeness based on a monte-carlo simulation of our survey for each population in percent assuming shown diameter (D) or larger and asteroid type. In ``()'' is the computed upper limit on the population size based on our survey coverage and null result.}
\end{deluxetable}
\end{center}

%
%
%
%

\begin{center}
\begin{deluxetable}{lccc}
\tablenum{4}
\tablewidth{6 in}
\tablecaption{Unstable Venus Co-orbital Completeness Percentage and Population}
\tablecolumns{4}
\tablehead{
\colhead{Type} & \colhead{D=0.5} & \colhead{D=1.0} & \colhead{D=1.5} \\ \colhead{} & \colhead{(km)} & \colhead{(km)} & \colhead{(km)}}  
\startdata
S-type  &  8\%($10^{+16}_{-7}$)   & 10\%($8^{+13}_{-6}$) & 10\%($8^{+13}_{-6}$)  \\
M-Type  &  6\%($12^{+20}_{-9}$)   & 10\%($8^{+13}_{-6}$) & 10\%($8^{+13}_{-6}$)  \\
E-Type  &  10\%($8^{+13}_{-6}$)   & 10\%($8^{+13}_{-6}$) & 10\%($8^{+13}_{-6}$)  \\
C-Type  &  0.4\%($220^{+360}_{-170}$)   & 7\%($11^{+17}_{-7}$)  & 10\%($8^{+13}_{-6}$)  \\
P-Type  &  0.2\%($440^{+720}_{-330}$)   & 5\%($14^{+23}_{-10}$)  & 9\%($8^{+13}_{-6}$)   \\
D-Type  &  0.3\%($310^{+510}_{-230}$)   & 7\%($11^{+19}_{-8}$)  & 10\%($8^{+13}_{-6}$)  \\
\enddata
\tablenotetext{}{Survey completeness based on a monte-carlo simulation of our survey for each population in percent assuming shown diameter or larger and asteroid type. In ``()'' is the computed upper limit on the population size based on our survey coverage and null result.}
\end{deluxetable}
\end{center}


%
%
%
%

\begin{center}
\begin{deluxetable}{lccc}
\tablenum{5}
\tablewidth{6 in}
\tablecaption{Atira/Apohele Asteroid Population Completeness Percentage and Population}
\tablecolumns{4}
\tablehead{
\colhead{Type} & \colhead{D=0.5} & \colhead{D=1.0} & \colhead{D=1.5} \\ \colhead{} & \colhead{(km)} & \colhead{(km)} & \colhead{(km)}}  
\startdata
S-type  &  7\%($55\pm30$)          & 8\%($50\pm25$)    & 8\%($50\pm25$)  \\
M-Type  &  6\%($65\pm35$)          & 8\%($50\pm25$)    & 8\%($50\pm25$)   \\
E-Type  &  8\%($50\pm25$)          & 8\%($50\pm25$)    & 8\%($50\pm25$)   \\
C-Type  &  0.4\%($950\pm500$)        & 7\%($60\pm30$)    & 8\%($50\pm25$)   \\
P-Type  &  0.3\%($1500\pm800$)        & 6\%($75\pm35$)    & 8\%($50\pm25$)   \\
D-Type  &  0.3\%($1200\pm600$)        & 6\%($65\pm30$)    & 8\%($50\pm25$)   \\
\enddata
\tablenotetext{}{Survey completeness based on a monte-carlo simulation of our survey for each population in percent assuming shown diameter (D) or larger and asteroid type. In ``()'' is the computed upper limit on the population size based on our survey coverage and finding 4 Atira asteroids within the survey.}
\end{deluxetable}
\end{center}

%
%
%
%

\begin{center}
\begin{deluxetable}{lccc}
\tablenum{6}
\tablewidth{6 in}
\tablecaption{Vatira Asteroids like 2020 AV2 Population Completeness Percentage}
\tablecolumns{4}
\tablehead{
\colhead{Type} & \colhead{D=0.5} & \colhead{D=1.0} & \colhead{D=1.5} \\ \colhead{} & \colhead{(km)} & \colhead{(km)} & \colhead{(km)}}  
\startdata
S-type  &  5\%($17^{+30}_{-13}$)   & 5\%($16^{+30}_{-12}$) &   5\%($16^{+30}_{-12}$)   \\
M-Type  &  5\%($19^{+30}_{-15}$)   & 5\%($16^{+30}_{-12}$) &   5\%($16^{+30}_{-12}$)  \\
E-Type  &  5\%($16^{+30}_{-12}$)   & 5\%($16^{+30}_{-12}$) &   5\%($16^{+30}_{-12}$)  \\
C-Type  &  0.7\%($130^{+220}_{-100}$)   & 5\%($18^{+30}_{-14}$)  &  5\%($16^{+30}_{-12}$)  \\
P-Type  &  0.5\%($170^{+270}_{-130}$)   & 5\%($19^{+30}_{-15}$)  &  5\%($16^{+30}_{-12}$)  \\
D-Type  &  0.5\%($160^{+270}_{-120}$)   & 5\%($19^{+30}_{-15}$)  &  5\%($16^{+30}_{-12}$)  \\
\enddata
\tablenotetext{}{Survey completeness based on a monte-carlo simulation of our survey for each population in percent assuming shown diameter (D) or larger and asteroid type. In ``()'' is the computed upper limit on the population size based on our survey coverage and null result.}
\end{deluxetable}
\end{center}
 
%
%
%
%

\begin{center}
\begin{deluxetable}{lccc}
\tablenum{7}
\tablewidth{6 in}
\tablecaption{Circular Vatira Asteroid Population Completeness Percentage}
\tablecolumns{4}
\tablehead{
\colhead{Type} & \colhead{D=0.5} & \colhead{D=1.0} & \colhead{D=1.5} \\ \colhead{} & \colhead{(km)} & \colhead{(km)} & \colhead{(km)}}  
\startdata
S-type  &  1\%   & 1\%    & 1\% \\
M-Type  &  1\%   & 1\%    & 1\% \\
E-Type  &  1\%   & 1\%    & 1\% \\
C-Type  &  0.3\% & 0.9\%  & 1\% \\
P-Type  &  0.3\% & 0.9\%  & 1\% \\
D-Type  &  0.3\% & 0.9\%  & 1\%  \\
\enddata
\tablenotetext{}{Survey completeness based on a monte-carlo simulation of our survey for each population in percent assuming shown diameter (D) or larger and asteroid type.}
\end{deluxetable}
\end{center}

\newpage

\newpage

\newpage

\newpage

\newpage

\newpage

\newpage

\newpage

\newpage

\end{document}